\documentclass[review]{elsarticle}

\usepackage{lineno,hyperref}
\usepackage{epstopdf}
\journal{Journal of \LaTeX\ Templates}

\begin{document}

\begin{frontmatter}

\title{Pd/Pt embedded CN monolayer as efficient catalysts for CO oxidation}

\author{Yong-Chao Rao}

\author{ Xiang-Mei Duan$^{\ast}$}
\address{Department of Physics, Faculty of Science, Ningbo University, Ningbo-315211, P.R. China}
\cortext[mycorrespondingauthor]{Corresponding author}
\ead{duanxiangmei@nbu.edu.cn}

\begin{abstract}
Single atom catalysts (SACs) based on 2D materials are identified to be efficient in many catalytic reaction. In this work, the catalytic performance of Pd/Pt embedded planar carbon nitride (CN) for CO oxidation, has been investigated via spin$-$polarized density functional theory calculations. We find that Pd/Pt can be firmly anchored in the porous CN monolayer due to the strong hybridization between Pd/Pt$-$$d$ orbitals and adjacent N$-$2$p$ orbitals. The resulting high adsorption energy and large diffusion barrier of Pd/Pt, ensuring the remarkable stability of the catalyst Pd/Pt@CN during CO oxidation reaction. The three distinct CO reaction mechanisms, namely, Eley$-$Rideal (ER), Langmuir$-$Hinshelwood (LH), and Tri$-$molecular Eley$-$Rideal (TER), are taken into consideration comparatively. Intriguingly, the oxidation reaction on Pd@CN prefers to proceed through the less common TER mechanism, where two CO molecules and one O$_{\rm 2}$ molecule need to cross a small reaction barrier of 0.48~eV, and finally dissociate into two CO$_{\rm 2}$ molecules. However, the LH mechanism is the most relevant one on Pt@CN with a rate$-$limiting reaction barrier of 0.68~eV. Moreover, the origin of SAC's reactivity enhancement is the electronic "acceptance$-$donation" interaction caused by orbital hybridization between Pd/Pt and preadsorbed O$_{\rm 2}$/CO.
Our findings are expected to widen the catalytic application of carbon based 2D materials.
\end{abstract}

\end{frontmatter}

\section{Introduction}
To control the emission of environmental$-$harmful pollutants and remove CO contaminants from H$_{\rm 2}$ gas fuel in fuel cells, the low$-$temperature CO oxidation reaction has received sustained attention.\cite{liu2012recent} Typical and conventional catalysts for CO oxidation mainly consisting of noble metals, especially for Pd\cite{nakai2006mechanism, gong2004systematic},
Pt\cite{eichler2002co, alavi1998co} and Au,\cite{liu2002catalytic, kimble2004reactivity} have been abroadly investigated for decades. However, the large$-$scale applications are restrained by the high cost and limited storage of these noble metals in earth.\cite{judai2004low, zhang2001density} Gradually accepted methods, downsizing the catalysts to atomic scale, for extreme case, SACs, can not only enhance the reactivity of catalysts, but also reduce the cost of fundamental scientific research and industrial applications.\cite{qiao2011single, moses2013co, thomas2005single} Meanwhile, it is an indisputable fact that high stability and no cluster formation of individual metal atom on the supporting material are the key prerequisites for catalyst to maintain its chemical activity.\cite{uzun2010site} Therefore research on finding an appropriate single-atom anchoring platform remains a major challenge.

Benefiting from the large surface area, high thermal stability, easily fabrication, two$-$dimensional (2D) materials, such as graphene\cite{krishnan2017new, li2010co} or defective graphene,\cite{esrafili2019exploring, zhang2015single},even Graphene assembled nanotube,\cite{Lu2017CO} Graphdiyne,\cite{Xu2018First} Boron sheet, \cite{zhu2019single} MoS$_{\rm 2}$,\cite{du2015mos, Ma2015CO} h$-$BN\cite{Mao2014A, gao2013co} and graphic carbon nitride (CN,\cite{Li2016Graphitic} C$_{\rm 3}$N$_{\rm 4}$\cite{gao2016single, Zheng2017Molecule} and C$_{\rm 2}$N\cite{yu2018c, li2018cu, Ma20163}), have been considered as prominent supports to host single atoms in various catalytic reactions (CO oxidation, O$_{\rm 2}$ reduction/evolution, H$_{\rm 2}$ evolution, N$_{\rm 2}$ reduction, $etc$). In particular, a porous 2D framework containing nitrogen can provide rich electron pairs to capture metal ions in the ligand. More importantly, the uniform nitrogen coordinators provides unique identification of active sites. Among these nitrogen$-$doped 2D carbon networks, CN has triggered intensive research in He separation,\cite{rao2019theoretical} H$_{\rm 2}$ storage\cite{Chen2018A} and O$_{\rm 2}$ evolution reaction.\cite{li2018cu} Although intrinsic CN is inert in electrocatalytic processes, it is still attractive to study whether the transition metals$-$embedded CN system can exhibit promising catalytic performance with high stability and superior reactivity in the CO oxidation reaction.

In this work, we have systematically investigated the catalytic performance of two representative species, Pd and Pt, anchored CN systems in CO oxidation process by employing first$-$principles calculations. The results show that TER mechanism is the most preferable one for CO oxidation on Pd@CN with a rate$-$determining barrier of 0.46~eV, while Pt@CN prefers the LH mechanism with a barrier of 0.68~eV. Moreover, the activation nature of O$_{\rm 2}$ and CO are interpreted through electron "acceptance$-$donation" interaction between gas molecules and transition metals. We predict that the Pd/Pt embedded CN system is an efficient, low cost and stable SACs with potential application in CO oxidation.

\section{Computational details}

The spin$-$polarized first$-$principles calculations were conducted in Vienna {\it ab initio} simulation package (VASP).\cite{Kresse1996Efficiency} The Projected Augmented Wave (PAW) in conjunction with Generalized Gradient Approximation (GGA) in the framework of Perdew$-$Burke$-$Ernzerhof (PBE) form with van der Waals (vdW) of DFT$-$D2 scheme was used to describe the electron$-$exchange correction effect,\cite{Perdew1998Generalized, Grimme2010Semiempirical} and a cutoff energy of 500~eV was adopted. The geometries were optimized until the energy and force were converged to 10$^{\rm -4}$~eV and 10$^{\rm -2}$~eV/{\AA}. To avoid interaction between two adjacent images, the vacuum space was set to be 20~\AA. Also, The Monkhorst$-$Pack meshes of $3\times3\times1$ were used in sampling the Brillouin zone. The on$-$site coulomb interaction (U) of 4~eV and exchange interaction (J) of 1~eV were applied to describe partially filled $d$$-$orbitals by considering coulomb and exchange corrections.\cite{Zhou2011Magnetism} The reaction paths were investigated by the climbing nudged elastic band (CI$-$NEB) method.\cite{Henkelman2000A} To evaluate the thermodynamic stability of Pd and Pt anchored CN systems, we carried out the first$-$principles molecular dynamics (MD) simulations at 300~K with a time step of 1~fs. Bader charge analysis\cite{Henkelman2006A} was used to obtain the amount of transferred electrons.

\section{Results and discussion}
\subsection{Structure and Stability of Pd/Pt embedded CN}

The optimized lattice parameter, corresponding C$-$N and C$-$C bond length of CN monolayer are 7.12, 1.34, and 1.51~\AA, respectively, coinciding with reported results.\cite{Li2016Graphitic, rao2019theoretical, Chen2018A} We used $2\times2$ CN supercell containing 24 carbon and 24 nitrogen atoms, creating a hole surrounding by six $sp{^{\rm 2}}$ bonded nitrogen atoms, which are beneficial for Pd/Pt anchoring. Taking Pd@CN as an example, it is indeed as expected that the Pd bonds to two N edge atoms with bond length of 2.20~{\AA} and keeps its planar geometric structure undistorted after structural optimization [shown in Fig. 1(a) and (b)]. Moreover, the energy band structures of for pure CN, Pd@CN and Pt@CN are presented in Fig. S1 (a) (b) (c), respectively. For Pd and Pt decorated CN, there are two bands dominated by N 2$p$ orbitals [verified by Fig. 1(e) and (f)] crossing the Fermi level, which results in a semiconductor to metal transition. The enhancement in conductivity makes Pd/Pt@CN promising for the electrocatalystic reaction of CO oxidation.
It is well known that transition metal (TM) atoms tend to form cluster on most of the substrate. The preference of clustering can be avoid when the electronic interaction between TM atoms and the substrate is sufficiently strong, and the mobility of TM atoms is low.
We first check the possibility of Pd and Pt aggregation on CN sheet. The binding energy ($E_{\rm b}$) of Pd/Pt on the CN monolayer is defined by $E_{\rm b}$ = $E_{\rm {Pd/Pt@CN}}$ $-$ $E_{\rm {Pd/Pt}}$ $-$ $E_{\rm CN}$, where $E_{\rm {Pd/Pt@CN}}$ is the total energy of Pd/Pt@CN, $E_{\rm {Pd/Pt}}$ and $E_{\rm CN}$ are the energies of spin$-$polarized Pd/Pt atoms and pure CN sheet, respectively. The considerably large $E_{\rm b}$ of $-$3.35~eV and $-$3.66~eV for Pd and Pt confirm that the metal atoms robustly bind with the adjacent N atoms. Besides, the relatively larger ratios, $E_{\rm b}$/$E_{\rm c}$ (cohesive energy, 0.85 for Pd, and 0.62 for Pt, respectively), further confirm that the two single TM atoms are not inclined to form cluster on CN.\cite{kittel2005introduction} The CI$-$NEB and MD methods were used to examine the dynamic and thermal stabilities of Pd/Pt@CN. As depicted in Fig. S2 (c), the large diffusion barriers, 2.88~eV and 2.45~eV for Pd and Pt, furthermore demonstrate that it is hard for single Pd/Pt to move on CN. MD simulations at room temperature [showed in Fig. S2 (d) and (e)] confirm that Pd/Pt@CN can keep its structure intact during CO oxidation reaction.
The electronic structure analyses show that Pd and Pt donate charge mainly to the edged $sp{^{\rm 2}}$$-$bound N atoms, as shown in Fig. 1(c) and (d). There are strongly overlapped peaks between Pd$-$4$d$/Pt$-$5$d$ and N$-$2$p$ orbitals in the projected DOS profiles [see Fig. 1(e) and (f)]. Especially, a peak of localized Pt$-$5$d$ orbital is situated near the Fermi level, resulting in an increase in the chemical activity of Pt@CN toward O$_{\rm 2}$ and CO.

\begin{figure}[h]
 \centering
 \includegraphics[width=9cm]{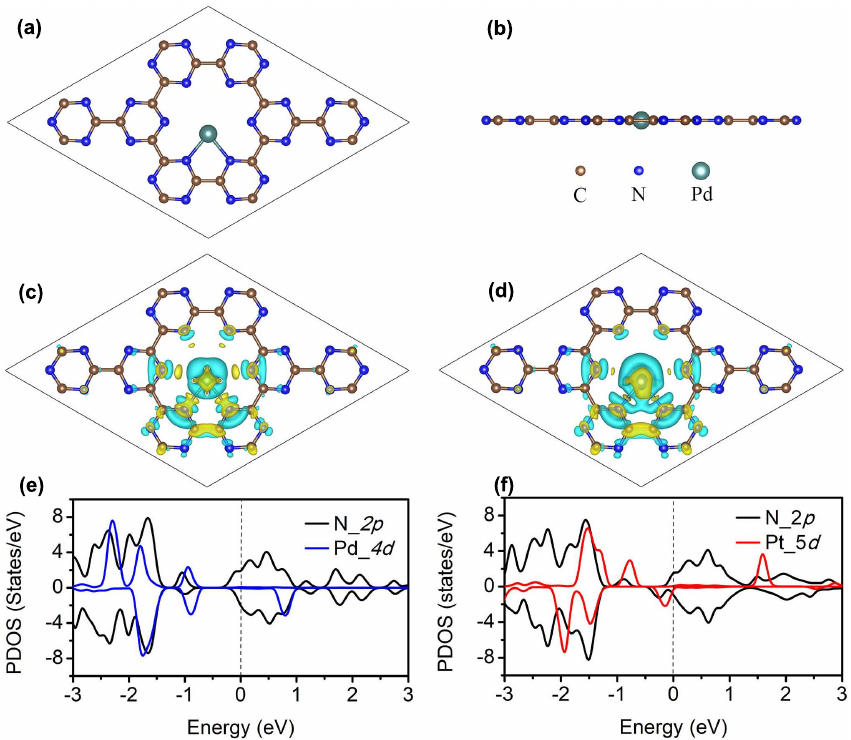}
 \caption{The top (a) and side view (b) of optimized atomic configurations for Pd@CN. The charge density difference plots (c) and (d) with an iso$-$surface value of 0.002~$e$/bohr${^{3}}$, as well as spin$-$polarized PDOS (e) and (f) for Pd@CN and Pt@CN, respectively. The blue and yellow iso-surfaces represent the electron depletion and accumulation, and the Fermi level is set zero.  }
\end{figure}

\subsection{O$_{\rm 2}$ and CO adsorption on Pd/Pt@CN}

For SACs, the efficiency strongly depends on its ability to bind the gas molecules over catalytic active centers. So the preliminary investigation of O${\rm _2}$ and CO adsorption over active sites, here Pd and Pt, is meaningful and essential. The adsorption energy ($E_{\rm a}$) of a single molecule (O$_{\rm 2}$, CO and CO$_{\rm 2}$) and O$_{\rm 2}$/CO+CO co$-$adsorption on the substrate were obtained by $E_{\rm a}$ = $E_{\rm {molecule+Pd/Pt@CN}}$ $-$ $E_{\rm molecule}$ $-$ $E_{\rm {Pd/Pt@CN}}$, where $E_{\rm {molecule+Pd/Pt@CN}}$, $E_{\rm molecule}$, $E_{\rm {Pd/Pt@CN}}$ presents the total energies of hybrid structures and isolated parts, respectively. Since the similarity of molecules adsorption on both Pd/CN and Pt/CN, the activation mechanisms of O$_{\rm 2}$ and CO on Pd@CN are taken as examples to discuss and we do not illustrate more details about gas molecules adsorption on Pt@CN (See Fig. S3) in the following part.

\begin{figure}[h]
 \centering
 \includegraphics[width=9cm]{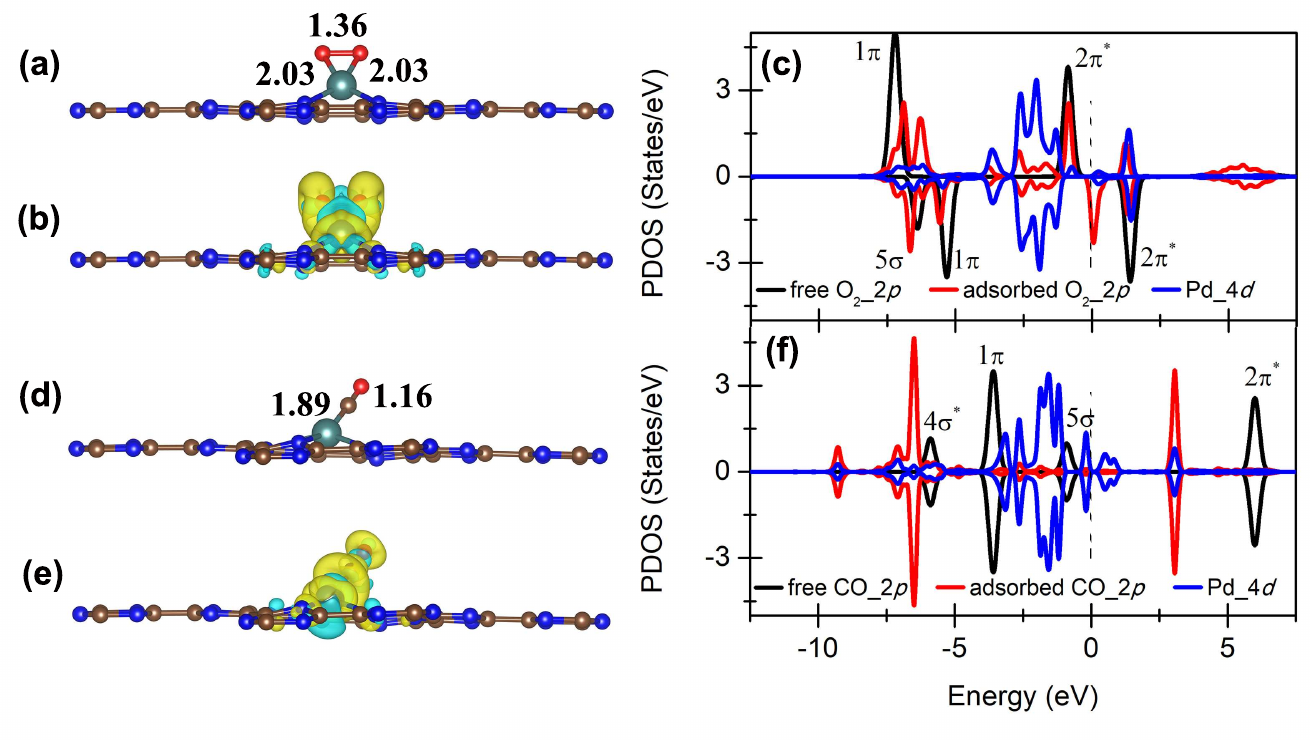}
 \caption{The side views of optimized structures and corresponding charge density difference plots with an iso$-$surface value of 0.002~{\it e}/bohr${^{3}}$, as well as spin$-$polarized PDOS for (a) (b) (c) O$_{\rm 2}$ and (d) (e) (f) CO adsorption on Pd@CN, respectively. The blue and yellow iso-surfaces represent the electron depletion and accumulation, and the Fermi level is set zero. }
\end{figure}

As shown in Fig. 2 (a), the O${\rm _2}$ molecule prefers to lie and parallel to the plane of CN sheet by side$-$on configuration, with the O$-$O bond length of 1.36~\AA, yielding a moderate adsorption energy of $-$0.69~eV. For comparison, the end$-$on configuration was taken into account, while the O${\rm _2}$ molecule will recline into the side$-$on one during the structural optimization. An general consequence of O${\rm _2}$ adsorption on the Pd@CN is that the distance of O$-$O bond is stretched from 1.23~{\AA} (in gas phase) to 1.36~\AA, indicating that the adsorbed O${\rm _2}$ is activated by the Pd@CN, which is of much significance for CO oxidation. Meanwhile, the Pd atom, originally locating in the undistorted plane of CN [shown in Fig. 1(b)], is also somehow pulled out from the sheet with two Pd$-$O distance of 2.03~{\AA} [see Fig. 2(a)]. Moreover, As depicted in Fig. 2(b), the charge accumulation region exhibits a feature of donuts surrounding two O atoms, and charge depletion is mainly around Pd, confirming that the charge transfer from Pd to O${\rm _2}$, which leads to the elongation of O$-$O bond. Bader analysis indicates that the O${\rm _2}$ molecule gains 0.48~$e$ from the substrate. For the two systems Pd@CN and Pt@CN (Fig. S3), the more electron transfer from the substrate to O${\rm _2}$, the larger elongation of O$-$O bond.

Since the reactivity of Pd@CN system originates from the activated O${\rm _2}$ adsorption state at the interface, so it is of significance to explore bonding characteristic between O${\rm _2}$ and Pd. As shown in Fig. 2(c), the O${\rm _2}$ molecule initially donates 1$\pi$ orbital electrons to the Pd$-$4$d$ orbital, and then the occupied Pd$-$4$d$ orbital interacts with the 2$\pi$$^{\rm \ast}$ orbital of O${\rm _2}$ molecule, causing a reverse charge transfer from Pd to O${\rm _2}$ molecule, and partially occupying the 2$\pi$$^{\rm \ast}$ orbital of O${\rm _2}$ (a spin$-$down peak occurs near the Fermi level). The accessibility of moderate electron donation and back$-$donation between adsorbed molecule and active site is highly desirable for the reactivity in most SACs, especially for the diatomic catalytic reactions, such as CO oxidation\cite{du2015mos} and N${\rm _2}$ reduction.\cite{zhu2019single, ling2018metal}

The most stable CO adsorption configuration as well as corresponding charge density difference plot, PDOS are presented in Fig. 2(d) (e) (f), respectively. The CO prefers an end$-$on configuration with the formation of a Pd$-$C bond (1.89~\AA). The adsorption energy ($E{\rm _a}$ = $-$0.97~eV) is moderate, which may avoid CO poisoning and surely benefits for the reaction. Similar to the adsorption of O${\rm _2}$, the Pd atom is also pulled out of the CN plane upon CO adsorption. The bond distance of adsorbed CO is slightly elongated by 0.02~{\AA} as compared with that of free CO (1.14~{\AA}), owing to the less electron transfer (CO gains 0.06~$e$). Combining the Fig. 2(e) and (f), it is found that the slight charge transfer, which is the net result of the donation of CO lone$-$pair electrons to Pd$-$4$d$ orbital and back$-$donation of Pd$-$4$d$ electrons to the CO$-$2$\pi$$^{\rm \ast}$ states.\cite{abild2007co}

Interestingly, the adsorption strength of O${\rm _2}$/CO well correlates with the electron transfer amount from the catalyst to the adsorbate when comparing the two systems [see Fig. 2 and Fig. S3, corresponding to the Pd@CN and Pt@CN, respectively]. The more electron transfer from Pt@CN to O${\rm _2}$/CO [0.6/0.08~$e$], the stronger adsorption strength of O${\rm _2}$/CO with Pt/CN substrate [$E{\rm _a}$(O${\rm _2}$/CO) = $-$1.45/$-$2.27~eV ].
The concern turns to that the high adsorption strength of CO molecule on Pt/CN inevitably leads to CO poison. Yet, the O${\rm _2}$$-$riched working environment is able to induce the Pt atom to bind with O${\rm _2}$ preferentially.\cite{liu2014co} In addition, the previous works concluded that the SACs, in which the adsorption strength of CO over active sites were in the same level with that of Pt/CN [$E{\rm_a}$(CO) = $-$2.39/$-$2.14~eV for Pt/MoS${\rm _2}$ \cite{du2015mos}and Cu${\rm _2}$/C${\rm _2}$N,\cite{li2018cu} respectively], possessed ideal efficiency for CO oxidation. Herein, Pt/CN is still a promising candidate for SACs in CO oxidation reaction.

We now consider the adsorption of atomic O and single CO${\rm _2}$ molecule over both Pd@CN and Pt@CN systems. As presented in Fig. S4, the distances of metal$-$O bond are all less than 2.0~\AA, indicate the stronger chemical interaction between Pd/Pt and O atom comparing with that of O${\rm _2}$, in line with the larger $E{\rm_a}$. The adsorption strength of CO${\rm _2}$ over Pd/Pt@CN substrate is still a key factor to determine whether these two systems could avoid CO${\rm _2}$ poison. Conventionally, the CO${\rm _2}$ adsorption energy should be less than 0.5~eV.\cite{Deng2013Single} It can be seen in the Fig. S4, the CO${\rm _2}$ adsorption energies on Pd/Pt@CN are $-$0.11 and $-$0.48~eV, respectively. Besides, the CO${\rm _2}$ molecule stays above the Pd/Pt@CN monolayer without any distortion, and the lengths of two C$-$O bonds are all equal to that of free CO${\rm _2}$ molecule. Therefore, we conclude that the molecule CO${\rm _2}$ can be released readily once formed.

Prior to study the reaction mechanism, it is necessary to check the capture performance (co$-$adsorption of the reactants) over the substrate. According to the Fig. S4, two reactants (O${\rm _2}$/CO and CO) are injected simultaneously into the Pd/Pt@CN, and form the co$-$adsorption configurations, either O${\rm _2}$+CO or CO+CO. The co$-$adsorption energies for these two configurations are larger than that of single O${\rm _2}$/CO molecule over both Pd@CN and Pt@CN. This indicates that the Pd/Pt@CN can host two reactants favorably. Herein, all above discussions demonstrate that Pd/Pt@CN can act as potent SACs for CO oxidation.

\subsection{CO Oxidation on Pd/Pt@CN}

Two well$-$established reaction mechanisms for CO oxidation over SACs, namely, ER (Eley$-$Rideal) and LH (Langmuir$-$Hinshelwood), as well as new TER (termolecular Eley$-$Rideal) are taken into consideration.\cite{Bleakley1999A, Deng2013Single, Mao2014A}

\begin{figure*}[!tp]
 \centering
 \includegraphics[width=12cm]{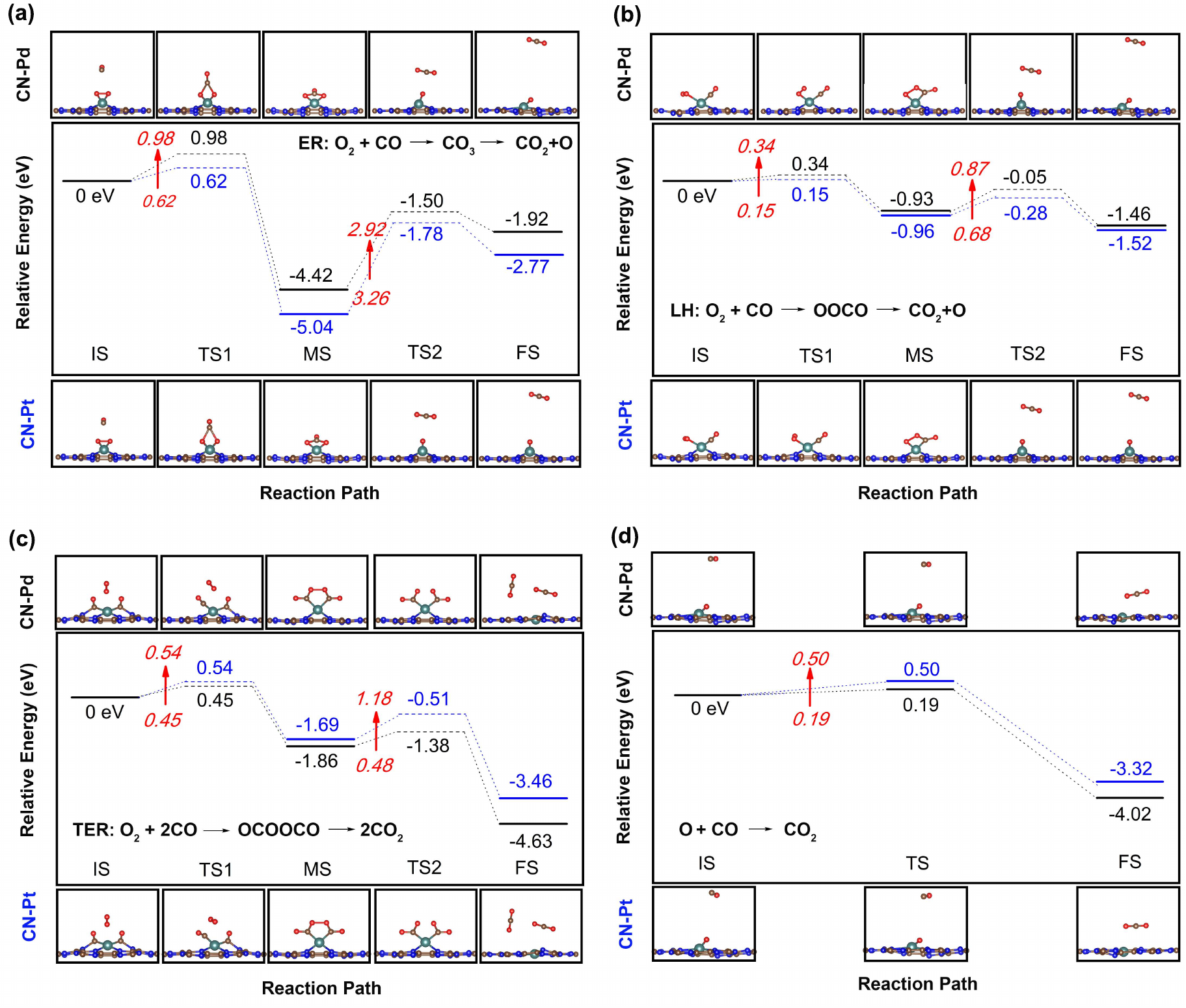}
 \caption{ Schematic energy profiles along the (a) ER, (b) LH, (c) TER mechanism, and (d) CO oxidation by atomic O along the minimum$-$energy pathway. All energies are given with respect to the reference energy (eV) of IS.}
\end{figure*}

According to the adsorption energies discussed above, although the Pd/Pt@CN sheets show stronger interaction with CO compared to O${\rm _2}$, the catalytic center can be pre$-$occupied by O${\rm _2}$ under O${\rm _2}$ rich working environment. So we investigate the oxidation of the first CO by pre$-$adsorbed O${\rm _2}$ along the ER path. Taking Pd@CN as an example, the configuration of physisorbed CO above the pre$-$adsorbed O${\rm _2}$ over Pd@CN was selected as initial state (IS)[see Fig. 3(a)], and a chemisorbed atomic O and a physisorbed CO${\rm _2}$ molecule as final state (FS).
After the optimization, the O$-$O bond length in the IS is elongated from 1.23~{\AA} {free molecule} to 1.37~{\AA}, and hence, the adsorbed O${\rm _2}$ is sufficiently activated, while the CO is located above substrate with a long distance of 2.92~{\AA} between C and O from O${\rm _2}$ and the adsorption energy of CO is 0.09~eV. In the first half reaction, CO binds with pre$-$adsorbed O${\rm _2}$ to form a carbonate$-$like intermediate state (MS, CO${\rm _3}$) with a high reaction barrier of 0.98~eV by passing over the transition state 1 (TS1). At TS1, the new C$-$O (here, O atom comes from O${\rm _2}$) starts to build and the O$-$O bond length is further stretched to 1.60~\AA. In the second half reaction, due to the breaking of the C$-$O and Pd$-$O bonds, passing over the transition state 2 (TS2) would experience a lager energy barrier of 2.92~eV, finally a CO${\rm _2}$ molecule forms, leaving an atomic O absorbed over Pd atom. Even though the small adsorption energy of CO${\rm _2}$ on atomic O adsorbed Pd/Pt@CN (0.10/0.42~eV) indicates that the formed CO${\rm _2}$ can be easily released at the room temperature, and the entire reaction process is exothermic. However, the ER mechanism for CO oxidation is prohibited because of the considerably large reaction barriers ($\sim$ 3~eV) for both Pd and Pt embedded CN monolayer.

We then examine the CO oxidation over Pd/Pt@CN through the LH mechanism by simultaneously adsorbing O${\rm _2}$ and CO. As mentioned before, the stronger ability for the co$-$adsorption of O${\rm _2}$ and CO demonstrates the feasibility of LH mechanism. Fig. 3(b) depicts the reaction path and energy profile for CO oxidation on Pd/Pt@CN. At the IS, both O${\rm _2}$ and CO molecules bind to the Pd/Pt atom with end$-$on configurations, where O${\rm _2}$ and CO are parallel and tilted to the CN sheet, respectively. As evidenced by the elongation of bond length in Fig. S4, reactants are slightly activated. And then one O atom (coming from O${\rm _2}$) without binding to Pd/Pt atom will approach to C atom of CO, forming a peroxo$-$type OOCO (MS) with the further enlargement of O$-$O bond (reaching up to 1.59~{\AA} for Pd and Pt@CN). In this first half process, passing over the TS1 needs to climb the energy barrier of 0.34~eV and 0.15~eV for Pd and Pt@CN, respectively. And in the second half process, the reaction continuously proceeds from MS to TS through TS2 with an energy barrier of 0.87 (Pd@CN) and 0.68~eV (Pt@CN). To break the O$-$O and Pd/Pt$-$C bonds, the value of energy barrier is much higher than that of the previous half reaction. Finally, a CO${\rm _2}$ molecule is formed, leaving an atomic O at the Pd/Pt. It was reported that the reaction could process readily with the barrier of 0.8~eV or less.\cite{Deng2013Single} Therefore, CO can be oxidized through the LH mechanism on Pt@CN with a relatively small energy barrier of 0.68~eV and a fully exothermic reaction energy of 1.52~eV.
The left O in O${\rm _2}$ will continue to react with the second incoming CO to form another CO${\rm _2}$. As Fig. 3(d) indicates, the configurations of CO staying above the atomic O on Pd/Pt@CN and CO${\rm _2}$ absorbed Pd/Pt@CN are set as IS and FS, respectively. For Pt@CN, the reaction barrier for the formation of second CO${\rm _2}$ is only 0.5~eV and the reaction releases a large heat of 3.32~eV. Therefore, the CO oxidation reaction over Pt@CN through LH mechanism is completely feasible.

A previous study has shown that two CO molecules can assist O${\rm _2}$ scission and promote CO oxidation.\cite{liu2013co} Moreover, the co$-$adsorption of two CO molecules over Pd/Pt@CN is stronger than that of single one. So the new TER route is worthy of consideration. The reaction route and relative energy are given in Fig. 3(c). Here, the CO oxidation reaction starts from the co$-$adsorption of two CO molecules over Pd/Pt (shown in Fig. S4). At the IS, the O${\rm _2}$ is physisorbed over the pre$-$adsorbed CO molecules with elongation of O$-$O bond to 1.26~A for Pd/Pt@CN. To process, O${\rm _2}$ is inserted into two CO molecules, leading to the formation of OCOOCO (MS) over Pd/Pt. The first half reaction would easily occur with a low energy barrier of 0.45 (Pd/CN) and 0.54~eV (Pt@CN). To pass over the TS2, the O$-$O bond length would be further extended until it breakes. After that, OCOOCO is decomposed into two free CO${\rm _2}$ molecules synchronously with a reaction barrier of 0.48 (Pd@CN) and 1.18~eV (Pt@CN). Overall, the rate$-$limiting reaction barrier for Pd@CN is only 0.48~eV, lower than those on noble metal catalysts (0.53$-$1.01~eV).\cite{gong2004systematic, eichler2002co} Additionally, the averaged adsorption energy for one CO${\rm _2}$ molecule in FS is 0.19~eV, which provides the possibility of spontaneous desorption from the catalyst Pd@CN.

\section{Conclusion}
In summary, by means of DFT calculations, we systemically investigate the potential of Pd and Pt anchored porous CN monolayer as SACs for CO oxidation. Both Pd and Pt possess appreciable binding energies to the CN monolayer along with large diffusion barriers, prohibiting the aggregation of adatoms, thereby enhancing the durability of catalysts. Besides, the MD simulation verifies that Pd/Pt@CN show highly thermodynamic stability at room temperature. Furthermore, the charge density difference plots and spin$-$polarized local density of state for O${\rm _2}$ and CO adsorption on Pd/Pt@CN demonstrate that the substrate can effectively capture and activate the gases via the electron "acceptance$-$donation" process. Finally, the detailed mechanistic pathways of CO oxidation over Pd/Pt@CN are comparatively investigated along ER, LH, and TER mechanisms. For Pd@CN, TER mechanism is more preferable, while Pt@CN is inclined to promote the CO oxidation through LH mechanism, and the rate$-$limiting energy barriers are less than those of noble metals catalysts. We predict that transition$-$metal atoms embedded in the porous CN monolayer is a promising platform to realize SACs for CO oxidation, which is worthy of further experimentation.

\section*{Acknowledgments}

This research is supported by the Natural Science Foundation of China (grant No. 11574167), the New Century 151 Talents Project of Zhejiang Province and the KC Wong Magna Foundation in Ningbo University.

\section*{Conflict of interest}
The authors declare they have no conflict of interest

\section*{References}

\bibliography{CN_SAC_for_CO_oxidation}

\end{document}